%% file: Main.tex
\documentclass[conference,compsoc]{IEEEtran}
\usepackage{subfigure}
\usepackage{amsmath}
\usepackage{fancyhdr}
\usepackage{multirow}
\usepackage{color}
\usepackage{MnSymbol}
\usepackage{mathrsfs}
\usepackage[table]{xcolor}
\usepackage{array}
\usepackage{placeins}
\usepackage{booktabs}

\usepackage{hyperref}
\hypersetup{
	colorlinks=false,
	pdfborder={0 0 0},
}

\usepackage{qtree}
\usepackage{tikz-qtree}

\usepackage[linesnumbered,ruled,vlined]{algorithm2e}

\SetCommentSty{mycommfont}

\SetKwInput{KwInput}{Input}                
\SetKwInput{KwOutput}{Output}              

\usepackage{lipsum}
\usepackage{makecell}

\usepackage{dblfloatfix}

\definecolor{bluegray}{rgb}{0.4, 0.6, 0.8}
\definecolor{orange1}{rgb}{1.0, 0.49, 0.0}
\definecolor{yellow1}{rgb}{0.99, 0.93, 0.0}
\definecolor{orange2}{rgb}{1.0, 0.75, 0.0}
\definecolor{green2}{rgb}{0.5, 1.0, 0.0}
\definecolor{green1}{rgb}{0.0, 0.5, 0.0}
\definecolor{AEC-green}{rgb}{0.47, 0.87, 0.47}
\definecolor{LA-blue}{rgb}{0.54, 0.81, 0.94}
\definecolor{darkgray}{rgb}{0.66, 0.66, 0.66}
\definecolor{blue1}{rgb}{0.19, 0.55, 0.91}
\definecolor{yellow2}{rgb}{1.0, 0.88, 0.21}
\definecolor{blush}{rgb}{0.87, 0.36, 0.51}
\definecolor{blue2}{rgb}{0.0, 0.75, 1.0}
\definecolor{green3}{rgb}{0.01, 0.75, 0.24}
\definecolor{red2}{rgb}{0.75, 0.0, 0.2}
\definecolor{byzantium}{rgb}{0.44, 0.16, 0.39} 
 	\definecolor{pansypurple}{rgb}{0.47, 0.09, 0.29}
	 	\definecolor{tyrianpurple}{rgb}{0.4, 0.01, 0.24}
 	\definecolor{aureolin}{rgb}{0.99, 0.93, 0.0}
\definecolor{americanrose}{rgb}{1.0, 0.01, 0.24}
\definecolor{burntorange}{rgb}{0.8, 0.33, 0.0}

\definecolor{cadmiumred}{rgb}{0.89, 0.0, 0.13}
\definecolor{cream}{rgb}{1.0, 0.99, 0.82}
\definecolor{champ}{rgb}{0.98, 0.84, 0.65}
\definecolor{black}{rgb}{0.0, 0.0, 0.0}
\definecolor{pink}{rgb}{1.0, 0.72, 0.77}
\definecolor{darkbrown}{rgb}{0.68, 0.33, 0.18}
\definecolor{darkpastelpurple}{rgb}{0.59, 0.44, 0.84}

\usepackage{graphicx}
\graphicspath{{Figures/}{Fig/}}
\DeclareGraphicsExtensions{.png,.PNG,.jpg,.jpeg,.pdf}

\renewcommand{\headrulewidth}{0pt}
\usepackage{lipsum}

\renewcommand\thefigure{\arabic{figure}}
\begin{document}

\title{IEA-Plugin: An AI Agent Reasoner for Test Data Analytics}
\author{
\IEEEauthorblockN{Seoyeon Kim, Yu Su, Li-C. Wang}
\IEEEauthorblockA{University of California, Santa Barbara\\
Santa Barbara, California 93106}
}

\maketitle

\begin{abstract}
This paper introduces IEA-plugin, a novel AI agent-based reasoning module developed as a new front-end for the Intelligent Engineering Assistant (IEA). The primary objective of IEA-plugin is to utilize the advanced reasoning and coding capabilities of Large Language Models (LLMs) to effectively address two critical practical challenges: capturing diverse engineering requirements and improving system scalability. Built on the LangGraph agentic programming platform, IEA-plugin is specifically tailored for industrial deployment and integration with backend test data analytics tools. Compared to the previously developed IEA-Plot (introduced two years ago), IEA-plugin represents a significant advancement, capitalizing on recent breakthroughs in LLMs to deliver capabilities that were previously unattainable.   
\end{abstract}


\thispagestyle{fancy}
\renewcommand{\headrulewidth}{0pt}
\fancyhf{}
\fancyfoot[R]{\large Submission}


%
\IEEEpeerreviewmaketitle


\vspace{-0.1cm}
\section{Introduction}
\vspace{-0.1cm}
\label{sec01}

\begin{figure}[htb]
	\centering
	\vspace{-0.2cm}
	\includegraphics[width=3.3in]{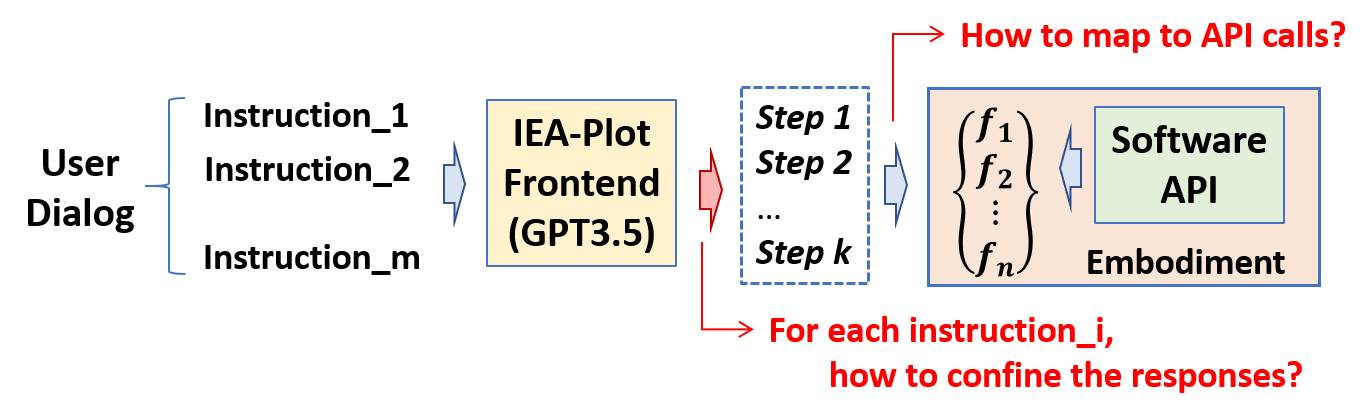}
	\vspace{-0.3cm}
	\caption{The task grounding problem considered in IEA-Plot in 2023 \cite{ITC2023}}
	\label{grounding}
	\vspace{-0.1cm}
\end{figure}

The introduction of IEA-Plot \cite{ITC2023} at ITC 2023 (where IEA stands for Intelligent Engineering Assistant) presented the design of an early AI agent tailored specifically for test data analytics. At that time, one of the most advanced models available was GPT-3.5 Turbo \cite{GPT-3.5-Turbo}. IEA-Plot leveraged the capabilities of GPT-3.5 Turbo, and was built upon its predecessor, IEA-2022 \cite{ITC2022}\cite{ITCAsia2022}, which utilized GPT-3 \cite{GPT3}. Unlike IEA-2022, which treated the mapping from user instructions to backend API calls as a language translation problem, IEA-Plot approached this mapping as a \textit{task grounding problem} \cite{Task-Grounding1}\cite{Task-Grounding2}, illustrated in Figure~\ref{grounding}. 

An input to IEA-Plot can be seen as a dialog comprising a sequence of user-provided instructions. At the backend, IEA-Plot interfaced with a software environment that provided various test data analytic tools. These tools were organized into an API supporting a set of functions, denoted as $f_1, \ldots, f_n$. A key constraint imposed by IEA-Plot was that the frontend's output had to be a sequence of executable steps, each realizable by one or a few of the available API functions. In most cases, the mappings between these steps and the available functions were one-to-one. Thus, the grounding problem involved correctly mapping each user instruction to one or few executable steps while ensuring that each step could be realized using a backend function. 

\begin{figure}[htb]
	\centering
	\vspace{-0.2cm}
	\includegraphics[width=3in]{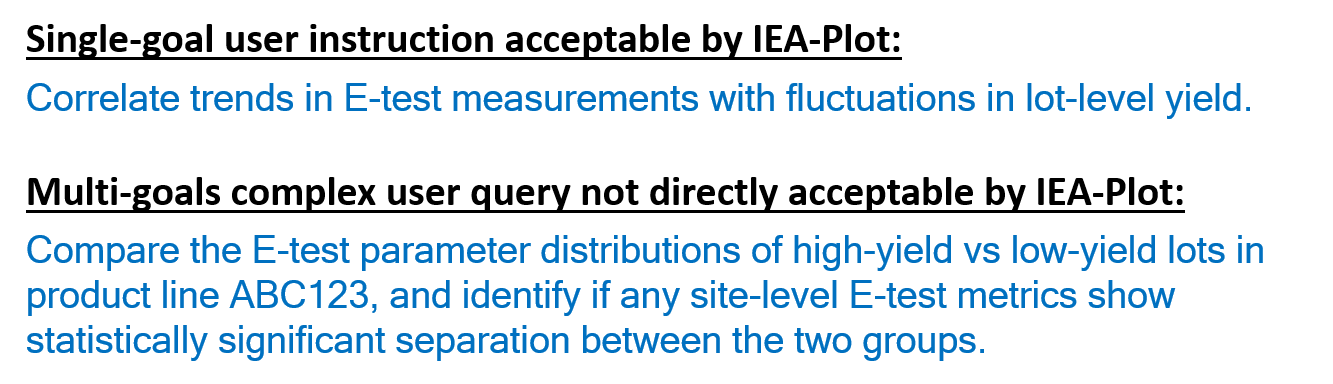}
	\vspace{-0.3cm}
	\caption{Single-goal instruction Vs. Complex multi-goals user query}
	\label{simplevscomplex}
	\vspace{-0.1cm}
\end{figure}

Treating the IEA-Plot frontend as solving a grounding problem implicitly imposed limitations on user input. Figure~\ref{simplevscomplex} illustrates this limitation. IEA-Plot assumed that a user instruction could be mapped to one (and occasionally few) predefined steps, each directly translated into a call to a specific function $f_i$ from the API. In Figure~\ref{simplevscomplex}, the first user input, representing a single-goal instruction, was therefore acceptable to IEA-Plot, provided there existed a function that implemented the correlation analysis between the average E-test value across all wafers from a lot and the lot yield. However, IEA-Plot would not be efficient for handling the second user input, representing a complex, multi-goal user query, as it required performing two single-goal tasks to fulfill the query: first performing a classification task to differentiate low-yield lots from high-yield lots, and then performing the requested correlation analysis. To support this query, a dedicated function connecting both tasks needed to be added to the backend of IEA-Plot.     

To overcome the deficiency, we sought to develop a new frontend capable of automatically generating a sequence of single-goal instructions --- referred to as a \textit{workflow} --- in response to complex user queries. Initially, this motivated us to design the new frontend which processed user queries before they reached the IEA-Plot.  

\vspace{-0.1cm}
\subsection{Two deployment obstacles faced by IEA-Plot}
\vspace{-0.2cm}
\label{sec01.1}

After the publication of IEA-Plot, we spent a year endeavoring to deploy IEA-Plot within a company. This experience exposed two significant challenges arising from the limitations of the IEA-Plot design: the first related to \textit{knowledge acquisition}, and the second concerning the \textit{scalability} of the IEA-Plot system.

\noindent
{\bf Knowledge Acquisition:} To tailor IEA-Plot to the specific needs of a team, it was natural to begin by gathering the team's requirements. Initially, we aimed to develop an exhaustive list of instructions that could serve as an {\it API specification} to guide backend tool development. However, this task proved quite challenging for two reasons.

Firstly, engineers and managers often could provide only a few examples or a general description of what they wanted IEA-Plot to accomplish, rather than a detailed list of specific instructions they anticipated using. Requesting a comprehensive list of instructions, required them to thoroughly document their current workflows and predict future needs--tasks that, in many cases, were simply impractical due to the substantial effort involved. Secondly, the impracticality was further exacerbated because the provided examples often turned out to be inherently more complex than the single-goal instructions considered by IEA-Plot (e.g. Figure~\ref{simplevscomplex}).

\noindent
{\bf Scalability:} Even if we had managed to develop a detailed specification of instructions for one team, the test organization encompassed multiple teams, each with unique requirements. If we implemented an API tailored to a specific team's specification, the tightly coupled design between user instructions and API functions in IEA-Plot meant that introducing new instructions from another team's specification could require substantial effort to modify the API structure. This lack of scalability significantly hindered our ability to efficiently accommodate diverse needs in the organization.  

\vspace{-0.1cm}
\subsection{Latest advancements in LLMs and AI}
\vspace{-0.2cm}
\label{sec01.2}

Recent advancements in Large Language Models (LLMs) have significantly improved their reasoning capabilities, enabling more effective handling of complex, multi-step problems. Techniques such as Chain-of-Thought (CoT) prompting \cite{wei2023chainofthoughtpromptingelicitsreasoning} have been instrumental by guiding models through explicit intermediate reasoning steps, as exemplified by Google's PaLM Model. A comprehensive survey by Plaat et al. \cite{plaat2024reasoninglargelanguagemodels} categorizes recent advances in LLM reasoning into prompting strategies, architectural enhancements, and novel learning paradigms. These combined innovations pave the way for more contextually intelligent AI applications.

GPT-o3-mini \cite{GPT-o3-mini}, released in January 2025, is a compact variant of OpenAI’s GPT-o3 model family, specifically optimized for efficient reasoning tasks. Despite its reduced parameter count, GPT-o3-mini maintains strong performance in logical inference, multi-step reasoning, and structured problem-solving. The availability of GPT-o3-mini and its variant GPT-o3-mini-high offered a promising new opportunity to addressing the challenges with IEA-Plot.

Additionally, LangGraph \cite{LangGraph_2024}, first released in stable form in June 2024, is an open-source Python library designed for building flexible and modular LLM-powered applications using computational graphs. It facilitates structuring complex AI workflows, known as {\it AI agents}, which involve multiple interactions among LLMs, integrated tools, and conditional logic within graph-based structures. These AI agents draw inspiration from recent agentic methods such as ReAct \cite{yao2023react} and Reflexion \cite{shinn2023reflexion}, and LangGraph significantly simplifies the implementation of them. 

Leveraging the capabilities of GPT-4o, GPT-o3-mini, GPT-o3-mini-high and LangGraph, in this work we share our experience developing the IEA-Plugin -- an AI agent that serves as a new frontend reasoner for IEA.

\subsection{IEA-Plugin: A new frontend reasoner}
\vspace{-0.2cm}
\label{sec01.3}

\begin{figure}[tb]
	\centering
	\vspace{-0.2cm}
	\includegraphics[width=3.1in]{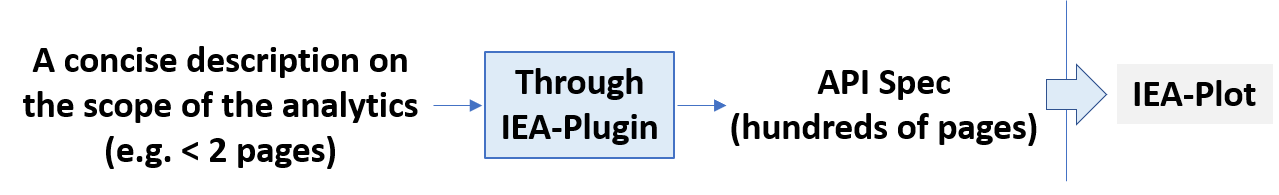}
	\vspace{-0.3cm}
	\caption{IEA-Plugin helps generate an API specification from a description}
	\label{IEAP-to-IEAP2}
	\vspace{-0.1cm}
\end{figure}

Figure~\ref{IEAP-to-IEAP2} shows the primary purpose of IEA-Plugin. One can start with a manually-created concise description on the scope of the analytics, e.g. less than 2 pages. It can be used to produce an API specification with hundreds of pages. This IEA-Plugin's API includes a list of {\it high-level} functions where each function, in view of the IEA-Plot depicted in Figure~\ref{grounding}, supports the execution of a {\it step}. When connecting to IEA-Plot, IEA-Plugin's functions would be realized by calling low-level functions in IEA-Plot, and if the functions were not available, they would need to be added.   

In operation, the primary usage of IEA-Plugin is to take a user query, which can be a complex, multi-goal query, as input and produce a {\it structured workflow} consisting of multiple instructions, which can then be realized by its high-level functions. Refer to Figure~\ref{grounding}: a {\it workflow} in IEA-Plugin's view corresponds to a {\it dialog} in IEA-Plot's view. 

Although IEA-Plugin was originally intended for IEA-Plot, after its completion we realized that effectively, IEA-Plugin could also operate independently. In standalone use, IEA-Plugin acts as a {\it knowledge acquisition} tool. Users interact with IEA-Plugin by submitting queries ranging from simple to complex. For each query, IEA-Plugin generates a workflow and stores the query-workflow pair in a centralized database. Users from different teams across the test organization can utilize IEA-Plugin to generate query-workflow pairs, all maintained within a single shared database. Subsequently, an LLM-assisted post-processing step distills these collected query-workflow pairs into structured API specifications for the backend of analytic tools. 

IEA-Plugin addresses the knowledge acquisition challenge by providing users with an interactive tool rather than directly requesting a complete specification upfront. By collecting various example queries—including complex, multi-goal queries—it systematically generates a comprehensive database of query-workflow pairs. Then, the API specification can be automatically distilled from this database. 

IEA-Plugin addresses the scalability challenge through a specification-distillation step performed after accumulating a substantial set of query-workflow pairs. Leveraging the summarization and coding capabilities of LLMs, it condenses a potentially large collection of instructions into a systematic, stable API structure. Consequently, although user queries can vary significantly in complexity and scope, the corresponding API structure required to support these diverse queries does not need frequent modification. In essence, IEA-Plugin effectively decouples user queries from the backend API structure by providing a stable framework upon which the backend can be efficiently expanded. 

The rest of the paper will detail the key innovations behind IEA-Plugin. 
Additional materials not shown in the paper can be accessed from the GitHub repository \cite{IEA-Plugin_2025}. 

\section{The Starting Phase}
\label{sec02}

\begin{figure}[htb]
	\centering
	\vspace{-0.4cm}
	\includegraphics[width=3in]{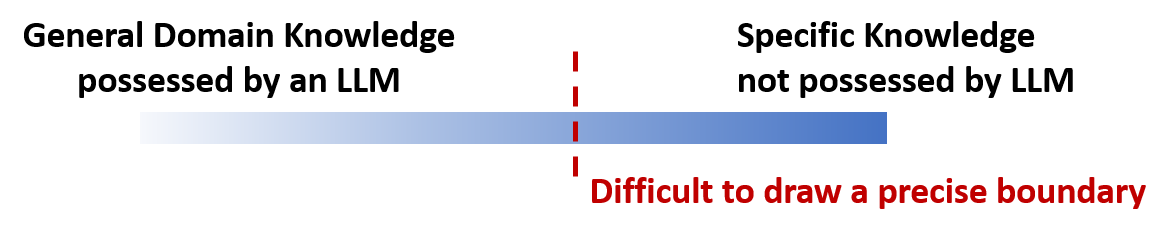}
	\vspace{-0.3cm}
	\caption{General domain knowledge Vs. Specific knowledge to a company}
	\label{knowledgedifficult}
	\vspace{-0.1cm}
\end{figure}

One intriguing question when utilizing an LLM to develop an AI application in a company-specific environment is related to drawing a boundary between the domain knowledge possessed by an LLM and the specific knowledge not possessed by the LLM. This is because if we knew that certain knowledge was not possessed by the LLM, then we would need to find a way to provide the knowledge. 

Note that assessing the knowledge content in LLMs has emerged as a growing field of research, with recent studies developing statistical approaches \cite{dong2023statisticalknowledgeassessmentlarge}, reliability metrics \cite{wang2023assessingreliabilitylargelanguage}, and systematic assessments using knowledge graphs \cite{luo2023systematicassessmentfactualknowledge}. Further, philosophical discussions emphasize challenges around evaluating intelligence and knowledge \cite{Bianchini2024EvaluatingIA}, while recent benchmarks examine knowledge boundaries through prompting techniques \cite{yin2024benchmarkingknowledgeboundarylarge}.

Our objective is not on assessing the knowledge content of an LLM. Rather, we desire to directly build an AI agent that can leverage the LLM's capabilities, including its knowledge in the domain of test data analytics. Nevertheless, it is important to keep in mind that an LLM might not have all the knowledge required to complete a given analytic task, especially when it involved knowledge from company-specific operation or infrastructure details.  

\begin{figure}[htb]
	\centering
	\vspace{-0.2cm}
	\includegraphics[width=3.3in]{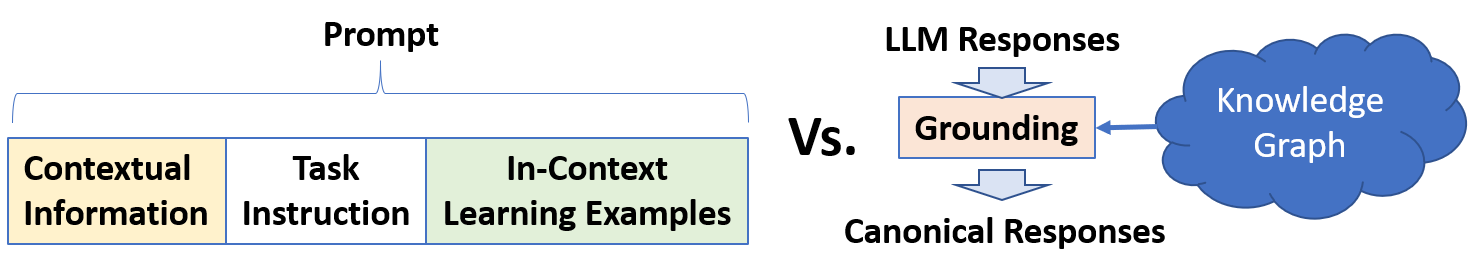}
	\vspace{-0.3cm}
	\caption{Activating knowledge Vs. Supplying knowledge}
	\label{supplyknowledge}
	\vspace{-0.1cm}
\end{figure}

Figure~\ref{supplyknowledge} illustrates two distinct ways to provide knowledge in an LLM application. Suppose the LLM possesses sufficient knowledge for a task. In this case, we use a prompt to {\it activate} the knowledge. The prompt can contain three separate portions: contextual information, task instruction, and in-context learning (ICL) examples. A prompt enables the LLM to perform either zero-shot learning (no ICL example) or few-shot learning (few ICL examples) \cite{GPT3}. This is in contrast to a traditional machine learning thinking where a large number of examples are provided for the model to learn the knowledge. 

Suppose the LLM does not have sufficient knowledge to properly perform a task. In this case, we can use a separate {\it grounding} approach for post-processing the LLM responses. This grounding can utilize the knowledge stored in a knowledge graph (KC) to convert LLM responses based on a set of acceptable and canonical responses prescribed by the KG. As a result, multiple LLM responses might be mapped into the same canonical response to remove their semantic variations. The design of IEA-Plot followed this approach, because at the time the GPT-3.5 model did not seem to have sufficient knowledge to do more. 

\subsection{An ICL-centric strategy}
\vspace{-0.2cm}
\label{sec02.1}

With the the availability of more advanced models (GPT-4o, GPT-o3-mini, etc.), there are opportunities to implement a new IEA component, based on not grounding but prompting. Since the release of GPT-3 \cite{GPT3}, including effective ICL examples in prompts has been recognized as a key strategy for unleashing the capabilities of LLMs. 

ICL examples can significantly enhance performance by providing explicit patterns for LLMs to generalize effectively without additional training. For instance, the authors in \cite{liu2021makesgoodincontextexamples} showed that semantically similar ICL examples improve task accuracy, while the authors in \cite{chen2024skillsincontextpromptingunlockingcompositionality} demonstrated that combining foundational and compositional skills in prompts unlocks advanced problem-solving capabilities. Additionally, the authors in \cite{wang2024knowledgeableincontexttuningexploring} highlighted the benefit of selecting knowledge-rich ICL examples to enhance factual correctness. These studies emphasize the crucial role of carefully chosen ICL examples for effective prompting. 

The development of the IEA-Plugin thus adopted an ICL-centric strategy. Considering the three components of a prompt shown in Figure~\ref{supplyknowledge}, our strategy emphasized the careful development and selection of ICL examples to enhance the performance of the LLM. The contextual information was fixed by providing a concise {\it scope description} (see Figure~\ref{IEAP-to-IEAP2}), and task instructions were usually kept concise, direct and intuitive, without excessive optimization.  

In our case, an ICL example is represented as a 2-tuple $(query, workflow)$, where $query$ is the input to the LLM-based workflow reasoner, and $workflow$ is the resulting output. Employing this ICL-centric strategy, the first step was to create an initial set of potential ICL examples.

\begin{figure}[hb]
	\centering
	\vspace{-0.3cm}
	\includegraphics[width=3.1in]{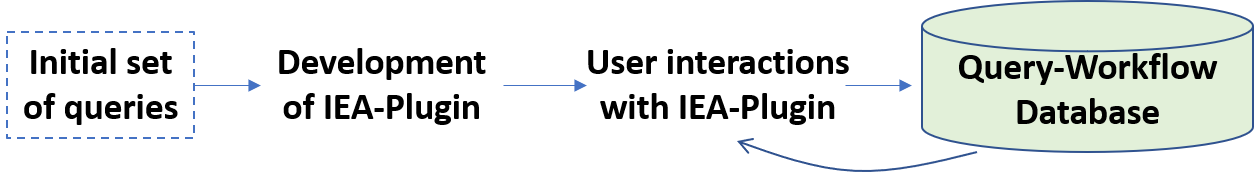}
	\vspace{-0.3cm}
	\caption{IEA-Plugin's operations for building a query-workflow database}
	\label{queryimportance}
	\vspace{-0.1cm}
\end{figure}

This initial set was stored in an example database, forming the foundation for processing subsequent queries. For example, upon receiving a new query, a small set of relevant examples from the database was selected as ICL examples for inclusion in the prompt. After processing, the newly generated query-workflow pair was added back into the database, expanding its content further. 

\subsection{Automatic query generation}
\vspace{-0.2cm}
\label{sec02.2}

\begin{figure}[htb]
	\centering
	\vspace{-0.2cm}
	\includegraphics[width=3.1in]{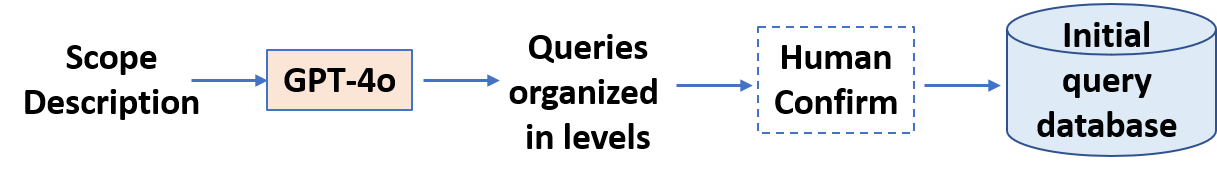}
	\vspace{-0.3cm}
	\caption{Obtaining an initial query database by leveraging GPT-4o model}
	\label{initialquery}
	\vspace{-0.2cm}
\end{figure}

To implement the approach depicted in Figure~\ref{queryimportance}, we needed to start with an initial set of queries. Figure~\ref{initialquery} shows our approach to automatically obtain this initial set, leveraging  capabilities of the GPT-4o model.

\begin{figure}[htb]
	\centering
	\vspace{-0.2cm}
	\includegraphics[width=3.3in]{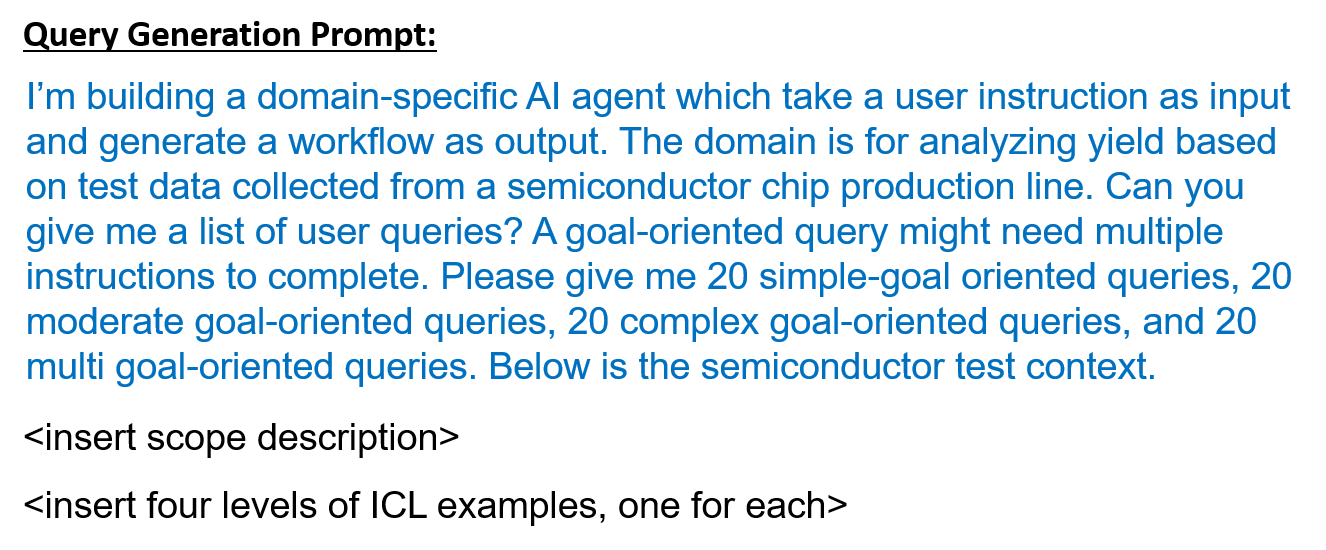}
	\vspace{-0.3cm}
	\caption{Prompt to generate initial query examples using GPT-4o model}
	\label{queryprompt}
	\vspace{-0.2cm}
\end{figure}

\begin{figure}[htb]
	\centering
	\vspace{-0.2cm}
	\includegraphics[width=3.3in]{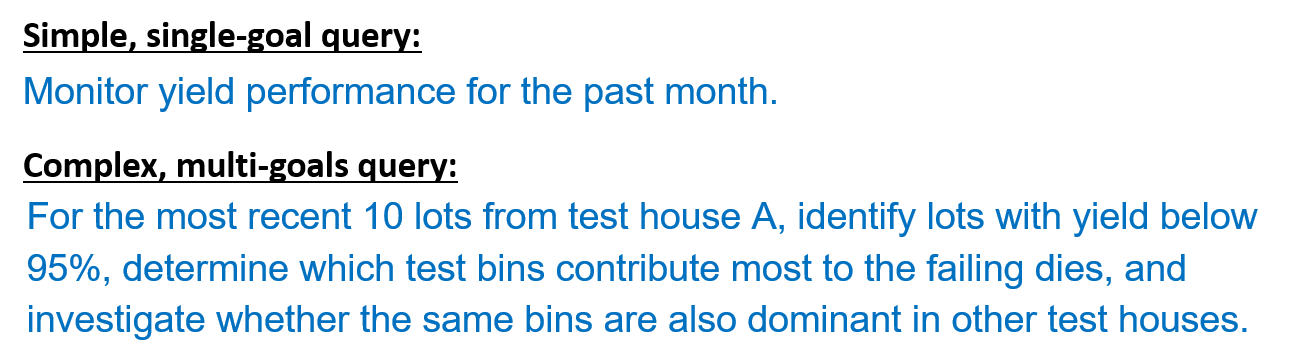}
	\vspace{-0.3cm}
	\caption{Examples of GPT-generated queries (see \cite{IEA-Plugin_2025} for more)}
	\label{queryexamples}
	\vspace{-0.1cm}
\end{figure}

Generation of the initial set of queries began with the scope description as shown in Figure~\ref{IEAP-to-IEAP2}. This description provided the application context for our query generation prompt, as illustrated in Figure~\ref{queryprompt}. The prompt requested queries at four complexity levels, ranging from simple to complex. The prompt included one ICL example per level. These initial ICL examples were selected from queries generated by the same prompt without any ICL examples. Figure~\ref{queryexamples} illustrates the quality of queries generated by GPT-4o through two examples at two levels. 

The automatic query-generation step serves two crucial purposes. First, manually creating a comprehensive set of queries covering diverse scenarios is challenging. In contrast, simply confirming and accepting queries generated automatically is significantly easier, thereby reducing the effort required to establish an initial query set. Second, this step also provides a means to assess the model’s domain knowledge. Through multiple iterations using varied scope descriptions, we observed that the GPT-4o model demonstrated a surprisingly strong understanding of semiconductor chip testing practices and test data analytics. 

Using the query generator, we obtained 80 queries, 20 at each level. Next we will discuss how to implement the reasoner to generate a workflow for each of the 80 queries and consequently, develop an initial query-workflow database as shown in Figure~\ref{queryimportance}.  

\section{The Workflow Reasoner}
\label{sec03}

\begin{figure}[htb]
	\centering
	\vspace{-0.2cm}
	\includegraphics[width=3.3in]{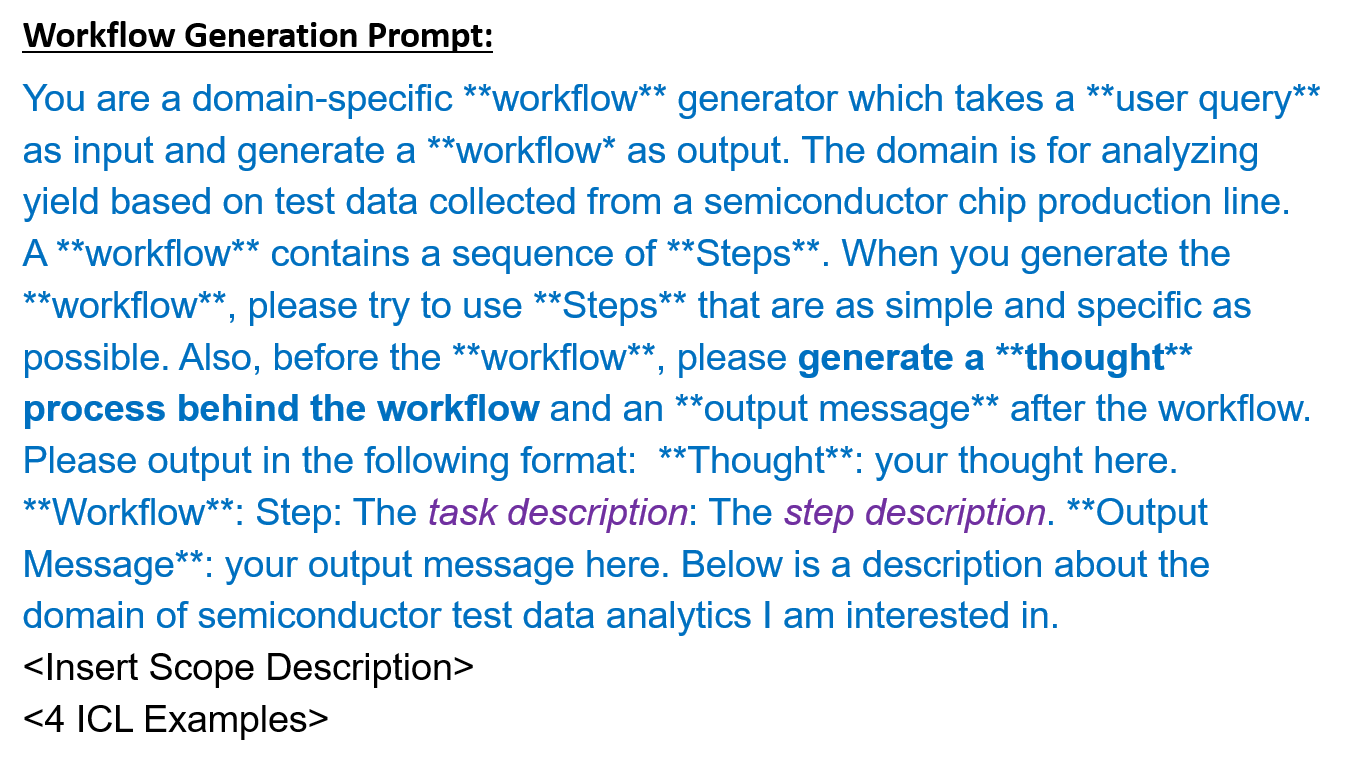}
	\vspace{-0.3cm}
	\caption{Workflow generation prompt with the scope of yield analysis}
	\label{WFprompt}
	\vspace{-0.2cm}
\end{figure}

The workflow reasoner was implemented as an AI agent using LangGraph Python library. A key step in the AI agent was to prompt the GPT-o3-mini model, the model designed specifically for reasoning tasks in the GPT family. Figure~\ref{WFprompt} shows the prompt. Prompting the model was done through the LLM calling interface in LangGraph. 

\subsection{Consistent ICL examples}
\vspace{-0.2cm}
\label{sec03.1}

The prompt included four ICL examples, initially consisting of two simple and two moderate single-goal-oriented queries (see Figure~\ref{queryprompt}). The first ICL example was obtained by applying the prompt (denoted as $P$) to the first simple query, without ICL examples. The response was reviewed, accepted, and added to $P$, creating $P[+1]$. We then applied $P[+1]$ to the second simple query, confirmed its response, and included it as the second ICL example, resulting in $P[+2]$. Repeating this process twice more produced $P[+4]$. 

This iterative approach ensured consistency among the four query-workflow pairs used as ICL examples, as each workflow was directly generated by the model without modification. Consequently, the primary purpose of the ICL examples was not to introduce additional domain knowledge but rather to instruct the model to "keep doing the same thing" when handling subsequent queries, thereby enhancing consistency in terms of workflow structure and wording. Similar strategies have been noted in prior research (e.g., \cite{liu2021makesgoodincontextexamples}), indicating that including semantically related ICL examples can improve the performance accuracy of LLMs. 

The prompt $P[+4]$ were then applied to the 20 simple queries and 20 moderate queries (including the original four) to generate 40 query-workflow pairs. These 40 pairs formed the initial database. 

\subsection{Including thought}
\vspace{-0.2cm}
\label{sec03.2}

The Chain-of-Thought (CoT) approach \cite{wei2023chainofthoughtpromptingelicitsreasoning} was among the earliest paradigms designed to enable reasoning. By explicitly prompting a model to articulate intermediate reasoning steps, the model could better perform complex reasoning tasks. As illustrated in Figure~\ref{WFprompt}, our prompt similarly instructs the model to ``generate a thought'' behind the workflow. Although GPT-o3-mini was developed for reasoning tasks and internally incorporates thought processes, we observed that explicitly including this instruction in the prompt could still significantly enhance its performance.

To demonstrate this, we modified the original prompt $P$ by removing the instruction of generating a thought, denoted as $P_{thoughtless}$. Following the procedure used to create the prompt $P[+4]$, we obtained a corresponding "thoughtless" prompt, $P_{thoughtless}[+4]$. We then applied $P_{thoughtless}[+4]$ to the 40 queries and compared the results against those generated with the prompt $P[+4]$.

For each query, we compared its two versions of the workflows by aligning their steps based on semantic similarities. Each step was represented as a semantic embedding vector of 1536 dimensions using the OpenAI {\it text-embedding-3-small} model. The semantic similarity between two steps was then computed using the cosine similarity of their embedding vectors. To identify the optimal matching between steps in two workflows, we utilized the Hungarian algorithm implemented in Scikit-Learn \cite{linearsum}. For each matched pair, the similarity score was computed, and then an average across all pairs was obtained, representing an overall {\it semantic similarity} measure between workflows. Steps without a matched counterpart were assigned similarity score 0.

\begin{figure}[htb]
	\centering
	\vspace{-0.2cm}
	\includegraphics[width=3.3in]{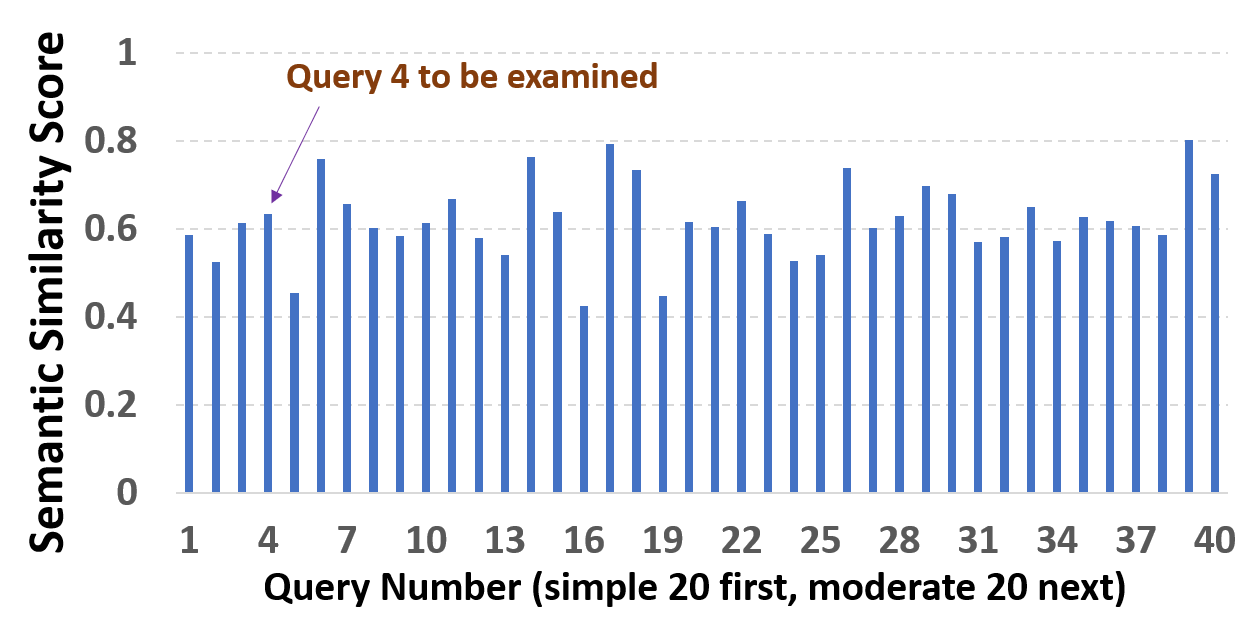}
	\vspace{-0.3cm}
	\caption{Semantic similarity scores between two versions of workflows}
	\label{thought}
	\vspace{-0.2cm}
\end{figure}

Figure~\ref{thought} presents the semantic similarity scores between each pair of workflows generated by $P[+4]$ and $P_{thoughtless}[+4]$. Most scores fall below 0.8, indicating the workflows differ significantly. Table~\ref{query4} illustrates these differences using query 4 as an example.  

\begin{table}[htb]
	\setlength{\tabcolsep}{2pt}
	\begin{tabular}{l|l} \toprule
	\multicolumn{2}{l}{Query: {\it List wafers with a consistent yield below 95\% over multiple weeks.}} \\
		Generated w/ thought & Generated w/o thought \\ \midrule 
		1. Define Weekly Intervals &  \\
		2. Retrieve Wafer Test Data & 1. Data Acquisition \\
									& 2. Data Cleaning \& Preparation \\
		3. Calculate Wafer-Level Yields & 3. Die-Level Pass/Fail Analysis \\
									& 4. Wafer Yield Calculation \\
									& 5. Temporal Aggregation \\ 
		4. Filter by Yield Threshold (95\%) &        \\        
		5. Identify Consistent Underperformers & 6. Consistency Analysis (95\%) \\
		6. Compile Results & 7. Result Listing \\
		7. Visualize Results & 8. Visualization \\
		                     & 9. Reporting \\ \bottomrule
	\end{tabular}
	\vspace{0.1cm}
	\caption{{\bf Query 4}: Two workflows (only task description of each step is shown) generated w/ and w/o thought}
	\label{query4}
\vspace{-0.2cm}
\end{table}

For simplicity, only the {\it task descriptions} for each step from both workflows are shown (see Figure~\ref{WFprompt} for the prompt requesting task description). A careful comparison reveals noticeable distinctions between the two workflows. The thoughtless prompt resulted in a more generic workflow. Note that steps 1–2 also appeared commonly across many workflows, and steps 3–4 frequently emerged in queries related to wafer yield. In contrast, the thoughtful prompt yielded a workflow tailored specifically to the query. With the thoughtful prompt, the model recognized the ambiguity of the term ``multiple weeks,'' prompting it to define this explicitly in the first step. Additionally, the model understood that identifying wafers ``consistently'' yielding below 95\% was crucial, leading step 4 to filter wafers based on this threshold and step 5 to address consistency explicitly. Moreover, the model treated the 95\% threshold as a key variable, handling it with a distinct step to allow the value to be changed in the future. Conversely, the thoughtless prompt combined both aspects into a single step (step 6).

Overall, we observed that the thoughtful prompt generated a workflow that avoided detailed data retrieval and calculation steps, in contrast to the workflow produced by the thoughtless prompt. The thoughtful prompt allocated additional steps specifically to the key aspects of the query (e.g., step 1 and steps 4–5), demonstrating that the model achieved a deeper understanding of the query.

\subsection{The AI agent}
\vspace{-0.2cm}
\label{sec03.3}

As mentioned earlier, the 40 query-workflow pairs formed our initial database. Recall that this database was accumulative, meaning its content could be expanded through usage. Beyond the original 40 queries, future queries were processed by selecting ICL examples from this database. This selection was performed by an AI agent that integrated the main ideas from both the RAG (retrieval-augmented generation) and ReAct \cite{yao2023react} agents.

\begin{figure}[htb]
	\centering
	\vspace{-0.2cm}
	\includegraphics[width=2.8in]{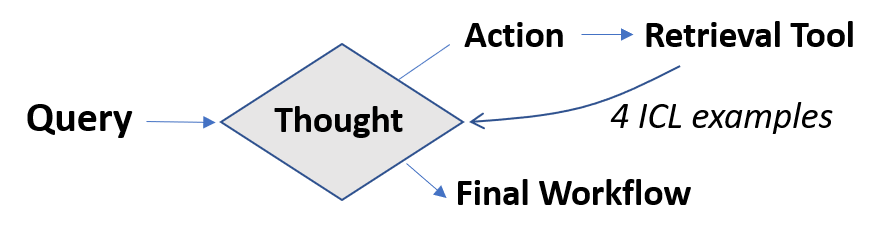}
	\vspace{-0.3cm}
	\caption{Our AI agent leveraging thought and retrieval action}
	\label{agent}
	\vspace{-0.1cm}
\end{figure}

Figure~\ref{agent} illustrates a high-level overview of our AI agent. When a query is received, the agent compares it with the queries stored in the database and retrieves the four queries that are semantically most similar. This is done using a standard similarity search approach that compares semantic embedding vectors—a method commonly used in RAG. These retrieved examples become the initial four ICL examples in the first iteration. Then, using the prompt in Figure~\ref{WFprompt} with these ICL examples, the agent generates a thought about the query. In the next iteration, retrieval is based on comparing this thought and thoughts from the database (instead of comparing queries themselves). This process repeats until the similarity score between two consecutively generated thoughts exceeds a threshold (e.g., 0.9). At that point, the agent ends its ``deliberation'' and outputs the workflow derived from the final thought.

Our agent combines the similarity search (retrieval-augmented) idea from RAG to select ICL examples and employs the iterative thought process from ReAct to determine deliberation reaching a convergent point. 

\begin{figure}[htb]
	\centering
	\vspace{-0.2cm}
	\includegraphics[width=3.3in]{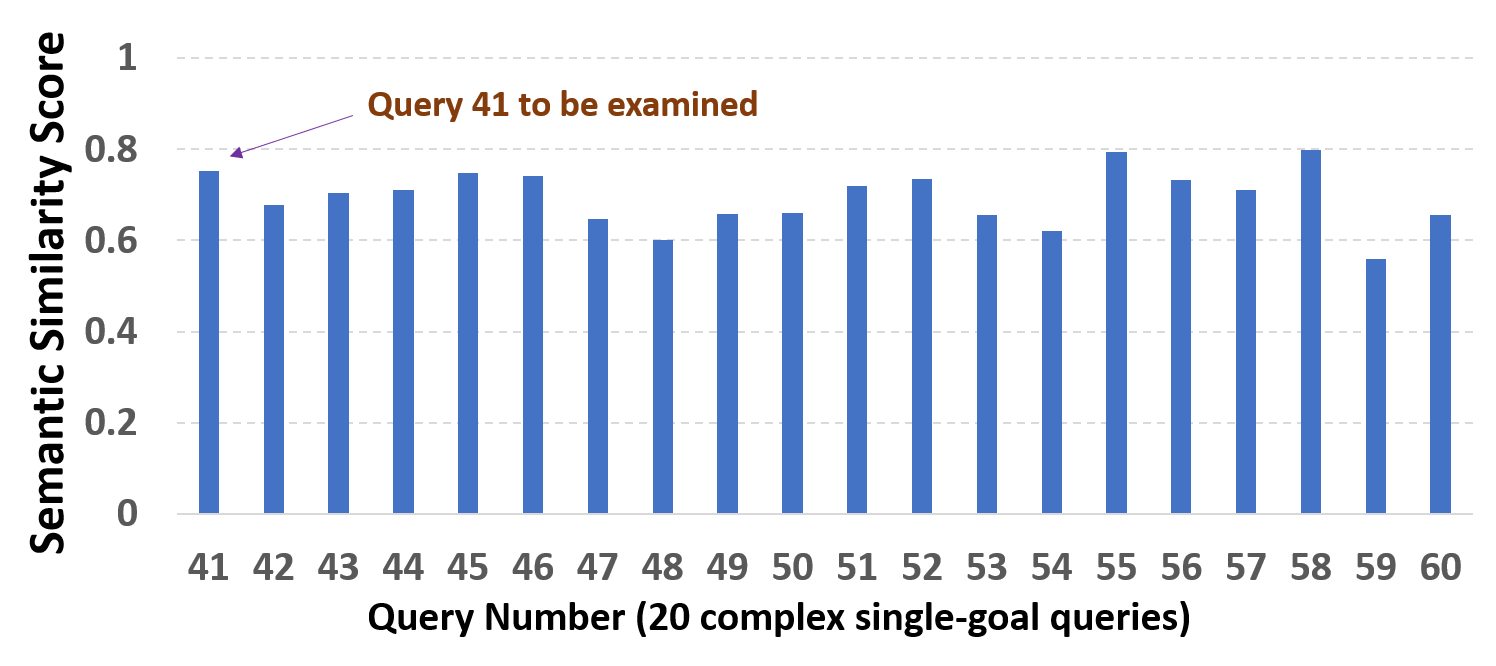}
	\vspace{-0.3cm}
	\caption{Semantic similarity scores between two versions of workflows}
	\label{react}
\end{figure}

Let $P_{agent}[+4]$ denote the final prompt used by the agent. Figure~\ref{react} compares two versions of the workflows generated by $P[+4]$ and $P_{agent}[+4]$ based on the 20 single-goal complex queries, following the same format as Figure~\ref{thought}. Most similarity scores fall below 0.8, indicating substantial differences between the two versions. Table~\ref{query41} illustrates an example of these differences using query 41.  

\begin{table}[htb]
	\setlength{\tabcolsep}{2pt}
	\begin{tabular}{l|l} \toprule 
		\multicolumn{2}{l}{Query: {\it Determine the correlation}} \\
		\multicolumn{2}{r}{\it between yield drops and specific manufacturing process changes.} \\
		Generated by the agent with $P_{agent}[+4]$ & Generated with $P[+4]$ \\ \midrule 
		1. Define Yield Drop Criteria &  \\
		2. Extract Yield Data & 1. Define Analysis Period \\
							  & 2. Extract Yield Data \\
							  & 3. Identify Yield Drops \\
		3. Identify Process Change Events & 4. Retrieve Process Change Data \\
		4. Align Data Temporally & 5. Align Timeframes \\
		5. Prepare Data for Analysis &  \\
		6. Perform Statistical Correlation Analysis & 6. Perform Correlation Analysis \\
		7. Evaluate Significance & \\
		8. Visualize the Results &   7. Visualize Findings \\         
		9. Generate Report and Insights & 8. Generate Analysis Report \\
    \bottomrule
	\end{tabular}
	\vspace{0.1cm}
	\caption{{\bf Query 41}: Two workflows (only task description of each step is shown) generated w/ and w/o the agent}
	\label{query41}
	\vspace{-0.4cm}
\end{table}

It is interesting to observe that the two workflows emphasized different aspects. The workflow generated by the agent focused on explicitly defining ``yield drop criteria'' and ``identifying process change events,'' whereas the workflow from $P[+4]$ emphasized ``defining the analysis period,'' by implicitly assuming that the criteria for determining yield drops were already established and that process change data was readily available. These distinctions highlight the impact of deliberation by the agent. A deeper examination of the semantics of the query suggests that the term "yield drops" required explicit definition, and the use of the word "specific" implied the existence of manufacturing process change data, consequently requiring identification of the changes that were correlated with the defined yield drops.

The agent's deliberated workflow differs from the $P[+4]$'s workflow in another key aspect: the correlation analysis. The agent explicitly divided this analysis into three distinct steps, recognizing that both data preparation and significance evaluation could independently affect the results. This separation is advantageous, as it directly influences the structure of the corresponding API. Having separate steps enhances flexibility, making it easier to accommodate various requirements for conducting the correlation analysis. 

\subsection{The LangGraph implementation} 
\vspace{-0.2cm}
\label{sec03.4}

\begin{figure}[htb]
	\centering
	\vspace{-0.2cm}
	\includegraphics[width=3.3in]{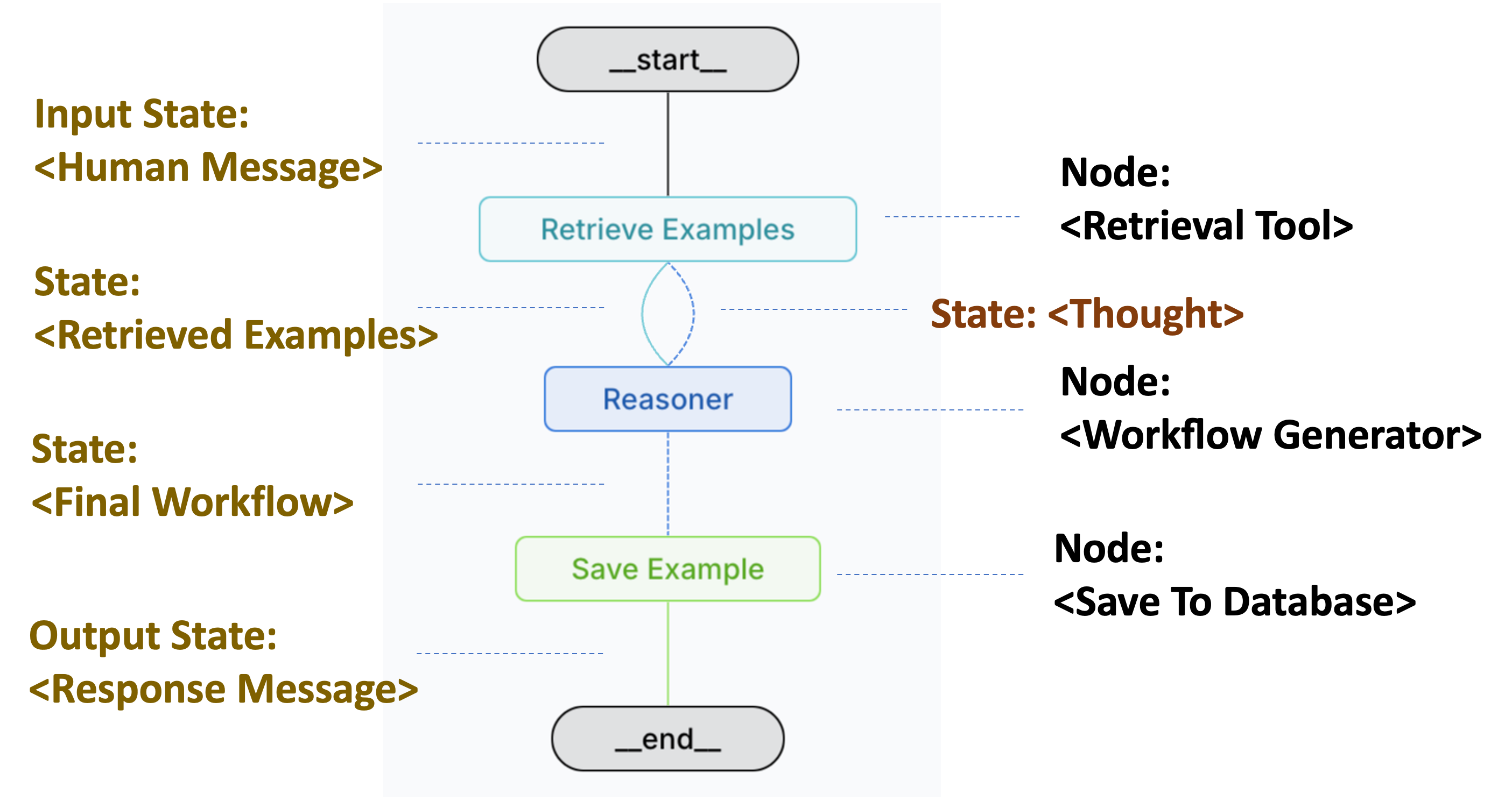}
	\vspace{-0.3cm}
	\caption{Screenshot of our AI agent shown in LangGraph Studio}
	\label{LG}
	\vspace{-0.3cm}
\end{figure}

Our AI agent reasoner was implemented using the LangGraph platform \cite{LangGraph_2024}. In LangGraph, an agent is modeled as an {\it executable} graph in which each {\it node} performs a specific function, such as calling an LLM or invoking a Python tool. Associated with each node are an {\it input state} and an {\it output state}. A {\it state} can be as simple as a dictionary, effectively a table of data. Figure~\ref{LG} presents a screenshot of our agent within the LangGraph Studio -- an interactive development environment integrated with LangGraph. 
 
The ``Reasoner'' node is where our agent invokes the GPT-o3-mini model. The ``Retrieve Examples'' node takes the thought generated by the ``Reasoner'' and conducts a similarity search against the query-workflow database. The retrieved examples allow the ``Reasoner'' to refine its thought process further. This iterative process continues until the reasoning stabilizes, as described earlier. At that point, the final workflow is saved to the query-workflow database via the ``Save Example'' node.

\subsection{Processing multi-goals queries}
\vspace{-0.2cm}
\label{sec03.5}

\begin{figure}[htb]
	\centering
	\vspace{-0.2cm}
	\includegraphics[width=3.3in]{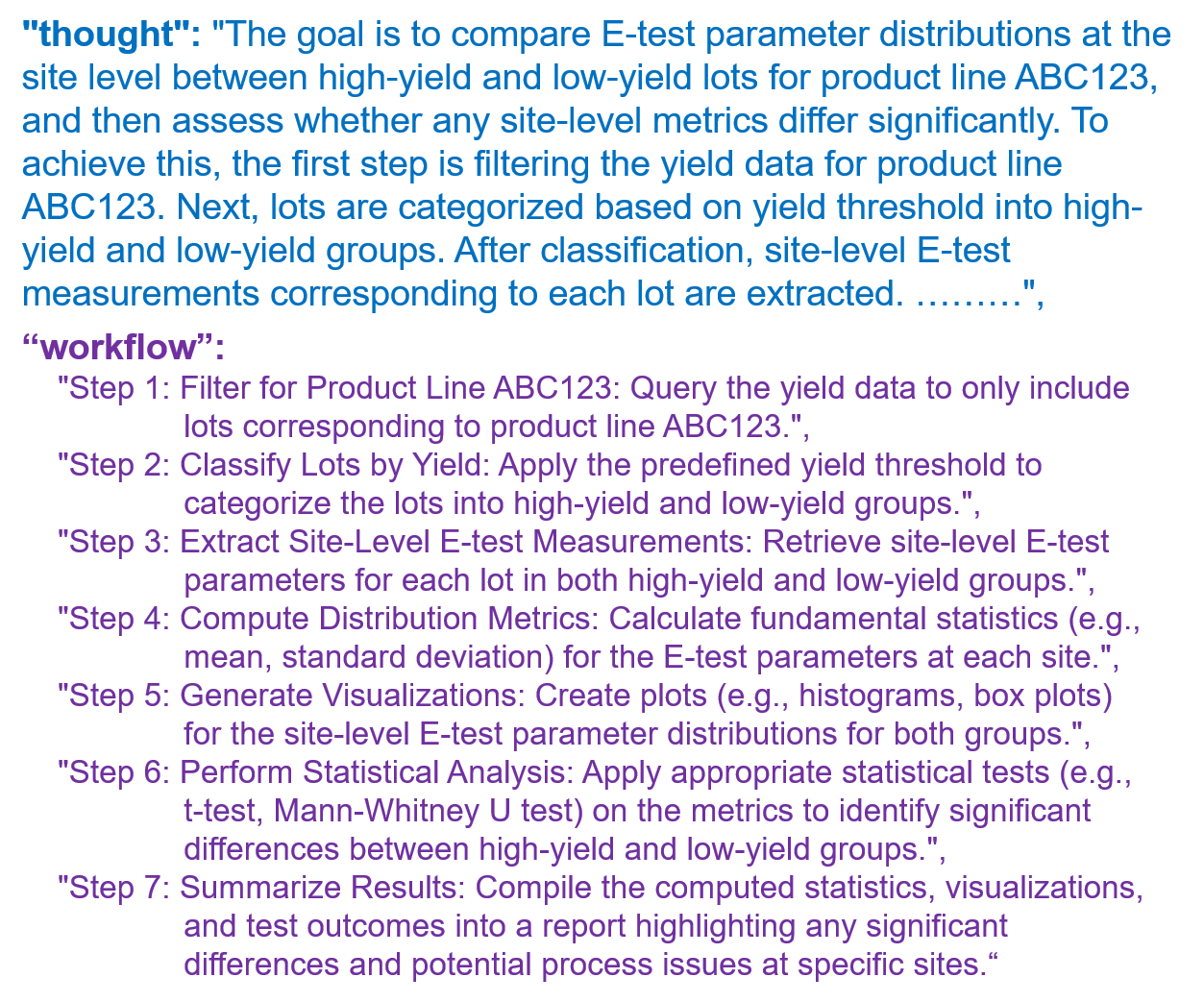}
	\vspace{-0.3cm}
	\caption{Output by the agent for the complex query in Figure~\ref{simplevscomplex}}
	\label{ComplexQ}
\end{figure}

After our agent processed the 20 complex single-goal queries and stored the resulting workflows in the database, we applied the agent to the remaining 20 multi-goal queries. Figure~\ref{ComplexQ} presents the agent's output for the multi-goal query initially introduced in Figure~\ref{simplevscomplex} before. 

As discussed in the Introduction section, this multi-goal query involves at least two objectives: classifying lots into low-yield versus high-yield categories, and correlating these classifications with E-test results. Interestingly, the generated workflow reveals that the model prioritized the primary goal -- the correlation analysis -- while condensing the secondary goal (classification) into a single step (Step 2). This automatic prioritization has implications when generating the API specification (see Section~\ref{sec04}), as the two steps will be treated differently. 

\subsection{There is no ``correct'' answer}
\vspace{-0.2cm}
\label{sec03.6}

Leveraging the reasoning capabilities of the GPT-o3-mini model, we demonstrated the feasibility of implementing an AI agent capable of taking a complex user query as input and producing a workflow. For professionals working in test and verification fields, a natural question arises: ``How do we know if the generated workflow is correct?'' However, this question can be problematic when evaluating such a workflow reasoner, for two primary reasons.

First, asking this question implies the existence of a definitive or objectively correct workflow. In practice, such an objective answer hardly exists. For instance, if the workflow is executed using IEA-Plot and produces plots, the user could review the results to determine whether the plots satisfy the intent expressed in the query. If unsatisfied, the user could simply refine or rephrase the query and repeat the process. This iterative approach is analogous to current interactions with LLMs through web interfaces, where correctness is  subjective -- dependent upon user preference rather than a fixed measure.

Second, LLM behavior is inherently statistical, meaning that identical prompts often yield varying workflow outputs. In our experience, this variability is especially prominent in reasoning-oriented models. Rather than viewing this statistical variability as a disadvantage, it can instead be advantageous. For example, IEA-Plugin could present multiple workflow variations, allowing users to select their preferred version. Here again, correctness becomes subjective (and probably more realistic) rather than absolute.

In our experiments, we consistently selected the first workflow generated by the model. To assess variability, we reviewed workflows from five repeated calls to the LLM for each query. We observed that the initial workflow produced was consistently the most intuitive. Statistically, the first response typically represents the ``most nominal'' workflow -- the one with the highest probability -- and, according to our observations, it was also consistently the most intuitive and appropriate version.

\section{Generation of API Specification}
\vspace{-0.2cm}
\label{sec04}

\begin{figure}[htb]
	\centering
	\vspace{-0.2cm}
	\includegraphics[width=3.3in]{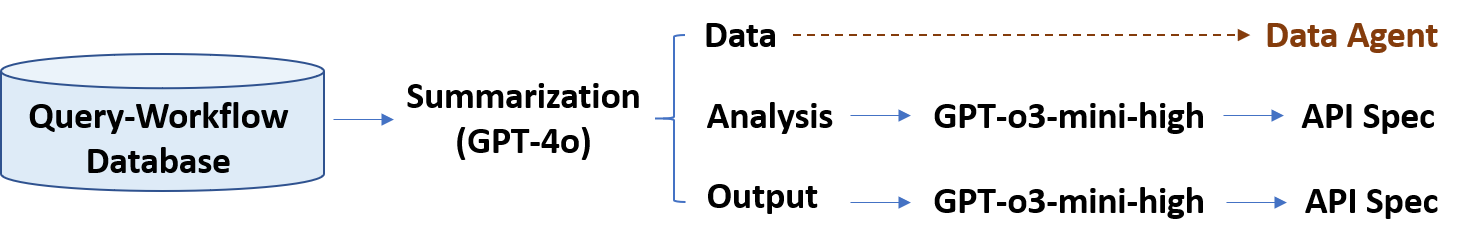}
	\vspace{-0.3cm}
	\caption{Illustration of our API specification generation process}
	\label{SpecG}
\end{figure}

After processing all 80 queries with our agent, we obtained a database containing 80 query-workflow pairs comprising a total of 588 steps. Using this database, we initiated an automated process to distill an API specification (Spec). Figure~\ref{SpecG} illustrates this process, which leverages the summarization capabilities of GPT-4o and the coding abilities of GPT-o3-mini-high. 

\subsection{Instruction Classification}
\vspace{-0.2cm}
\label{sec04.1}

For each step, we devised a prompt (see \cite{IEA-Plugin_2025}) instructing GPT-4o to extract representative terms for every workflow step. These terms are structured as a JSON-formatted list: $[``overall\_action'', ``action'', ``object'', ``attributes'']$. The first two terms ($overall\_action$ and $action$) are required to be single words, e.g. a verb. The difference between them is that $overall\_action$ is extracted from the {\it task description} of the step while $action$ extracted from the {\it step description} (see prompt in Figure~\ref{WFprompt}). The term $object$ indicates the entity upon which the action is performed, while $attributes$ are terms describing details associated with the $object$.

After extracting the terms for all steps, we used a second prompt instructing GPT-4o to classify each step into one of three categories: {\bf Analysis}, {\bf Output}, or {\bf Data}. Steps involving data retrieval and preparation were classified under the Data category. Steps related to result summarization and output generation were assigned to the Output category. Steps not in these two categories were classified as Analysis. 

\begin{table}[htb]
	\centering
	\begin{tabular}{ccc|ccc} \toprule
    \multicolumn{3}{c}{From original 80 queries} & \multicolumn{3}{c}{From 40 extra queries} \\ 
    {\bf Analysis} & {\bf Output} & {\bf Data} & {\bf Analysis} & {\bf Output} & {\bf Data} \\ \midrule
    199 &  135  & 254   & 112 & 63  &  133  \\ 
    \bottomrule
	\end{tabular}
	\vspace{0.1cm}
	\caption{Summary of categories for the 588 steps from the original 80 queries and for the 308 steps from the additional 40 queries, all classified by GPT-4o model. } 
	\label{category}
	\vspace{-0.2cm}
\end{table}

Table~\ref{category} summarizes the results of the classification performed by GPT-4o. Additionally, we instructed the model to generate 40 extra queries, following the approach described in Section~\ref{sec02.2}. These queries included 10 examples at each of the four complexity levels. We then used the AI agent to generate workflows for each query, resulting in a collection of 308 steps added to our database. Finally, we applied the same classification process to categorize these 308 new steps, and the results are also included in Table~\ref{category}. 

\subsection{Generation of API functions}
\vspace{-0.2cm}
\label{sec04.2}

Using the information obtained from the classification process, we devised a prompt (see \cite{IEA-Plugin_2025}) leveraging the coding capabilities of GPT-o3-mini-high to generate API specifications. As illustrated in Figure~\ref{SpecG}, API specifications were generated separately for steps categorized as Analysis and Output. We intentionally omitted generating an API specification for the Data category steps, as these would be automatically handled by a dedicated {\it Data} agent capable of generating code for database queries and basic calculations, which will be elaborated further in Section~\ref{sec04.2.3}.  

\subsubsection{Functions in the ``Analysis'' Category}
\label{sec04.2.1}

Table~\ref{AnalysisAPI} summarizes the results from the API generation process. Observe that for the original 80 queries, the model decided to use 141 API functions to implement the 199 steps. Then, by adding 112 steps from the extra 40 queries, only 47 functions were added. The ratio drop (from $\frac{141}{199}=70.85\%$ to 
$\frac{47}{112} = 41.96$) indicates that some functions for the original 199 steps were reused for the additional 112 steps. 

\begin{table}[htb]
	\centering
	\setlength{\tabcolsep}{2pt}
	\begin{tabular}{cc|cc} \toprule
		\multicolumn{2}{c}{From original 80 queries} & \multicolumn{2}{c}{From 40 additional queries} \\ 
		{\bf \# of Steps} & {\bf \# of Functions} & {\bf Added \# of Steps} & {\bf Added \# of Functions} \\ \midrule
		199 &  141  & 112 & 47  \\ 
		\bottomrule
	\end{tabular}
	\vspace{0.1cm}
	\caption{{\bf Analysis:} \# of API functions automatically derived from the original 80 queries and additional 40 queries} 
	\label{AnalysisAPI}
	\vspace{-0.4cm}
\end{table}

\begin{figure}[htb]
	\centering
	\vspace{-0.2cm}
	\includegraphics[width=3.3in]{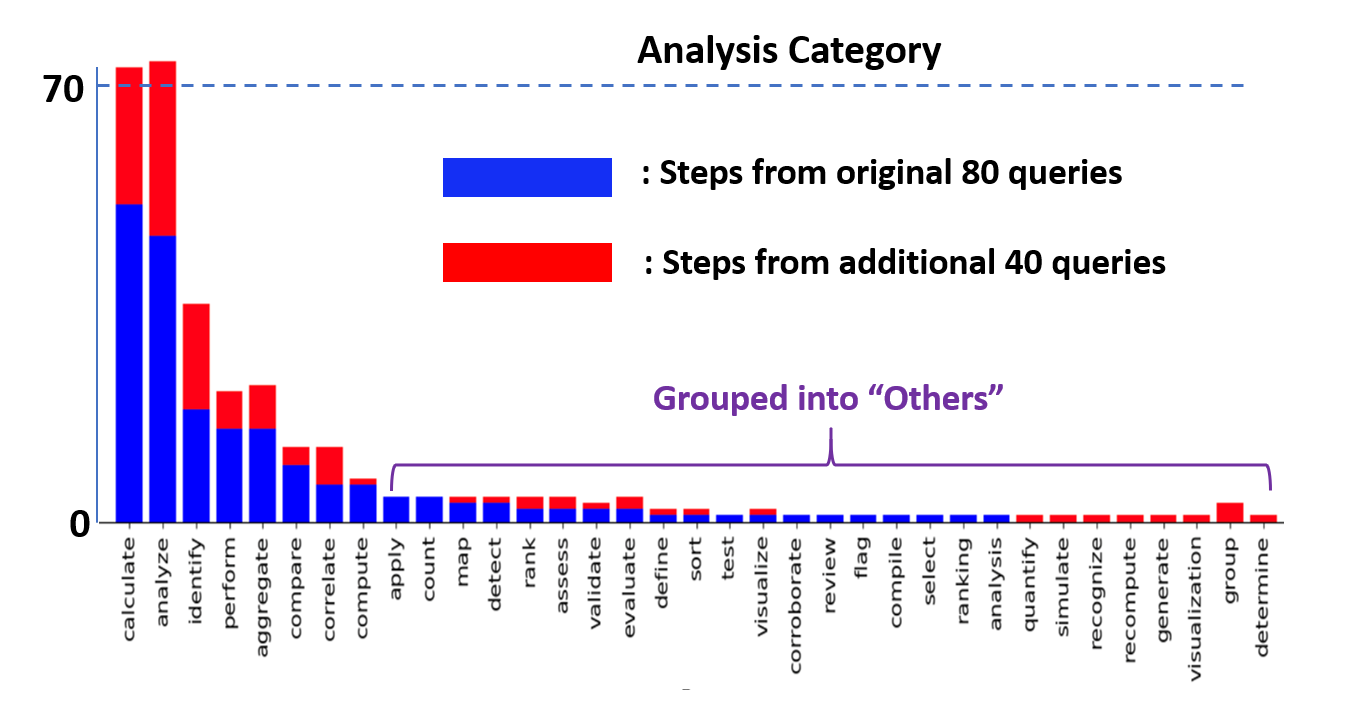}
	\vspace{-0.3cm}
	\caption{Histogram to show the \# of Steps classified into different actions}
	\label{AnalysisH1}
\end{figure}

Figure~\ref{AnalysisH1} presents a histogram detailing the distribution of steps across different $action$ groups. The two largest groups are labeled with actions {\tt analyze} and {\tt calculate}, followed by other significant groups including {\tt identify}, {\tt perform}, {\tt aggregate}, {\tt compare}, {\tt correlate}, among others. The figure compares the distribution of steps derived from the original 80 queries with those from the additional 40 queries. Notably, the extra 40 queries introduced several new $action$ labels (shown toward the right of the histogram), such as {\tt quantify}, {\tt simulate}, {\tt recompute}, {\tt generate}, etc. 

\begin{figure}[htb]
	\centering
	\vspace{-0.2cm}
	\includegraphics[width=2.4in]{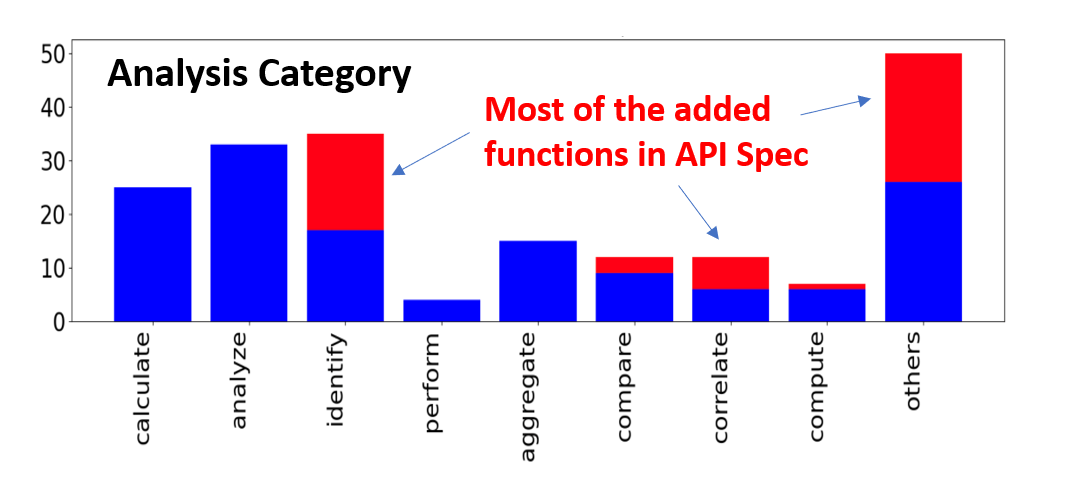}
	\vspace{-0.3cm}
	\caption{Histogram to show the \# of added functions in API Spec}
	\label{AnalysisH2}
\end{figure}

Figure~\ref{AnalysisH2} presents the corresponding histogram after grouping the API functions generated by the model. The largest 8 groups from Figure~\ref{AnalysisH1} are carried over into Figure~\ref{AnalysisH2}, while the remaining smaller groups from Figure~\ref{AnalysisH1} are consolidated into a single {\tt others} group. The generation of API functions was applied separately to each group of the  9 groups shown in Figure~\ref{AnalysisH2}, to ensure the number of steps remained within the input size limits of the GPT-o3-mini-high model, a constraint observed during our experiments. Referring to Table~\ref{AnalysisAPI}, the sum of all blue bars in Figure~\ref{AnalysisH2} is 141, and the sum of all red bars is 47. 

The API documentation generated by GPT-o3-mini-high can be found in the IEA-Plugin repository~\cite{IEA-Plugin_2025}. Figure~\ref{AnalysisH2} illustrates an interesting observation: functions belonging to the action groups {\tt calculate}, {\tt analyze}, {\tt perform}, and {\tt aggregate} were shared across steps from both the original 80 queries and the additional 40 queries. Conversely, steps in the action groups {\tt identify}, {\tt correlate}, and {\tt others} showed almost no overlap; those steps from the additional 40 queries predominantly introduced new functions. These findings highlight an essential characteristic of the API structure prescribed by the models, wherein reusable functions and those tailored to unique tasks are clearly separated into distinct groups. 

\begin{figure}[htb]
	\centering
	\vspace{-0.2cm}
	\includegraphics[width=3.3in]{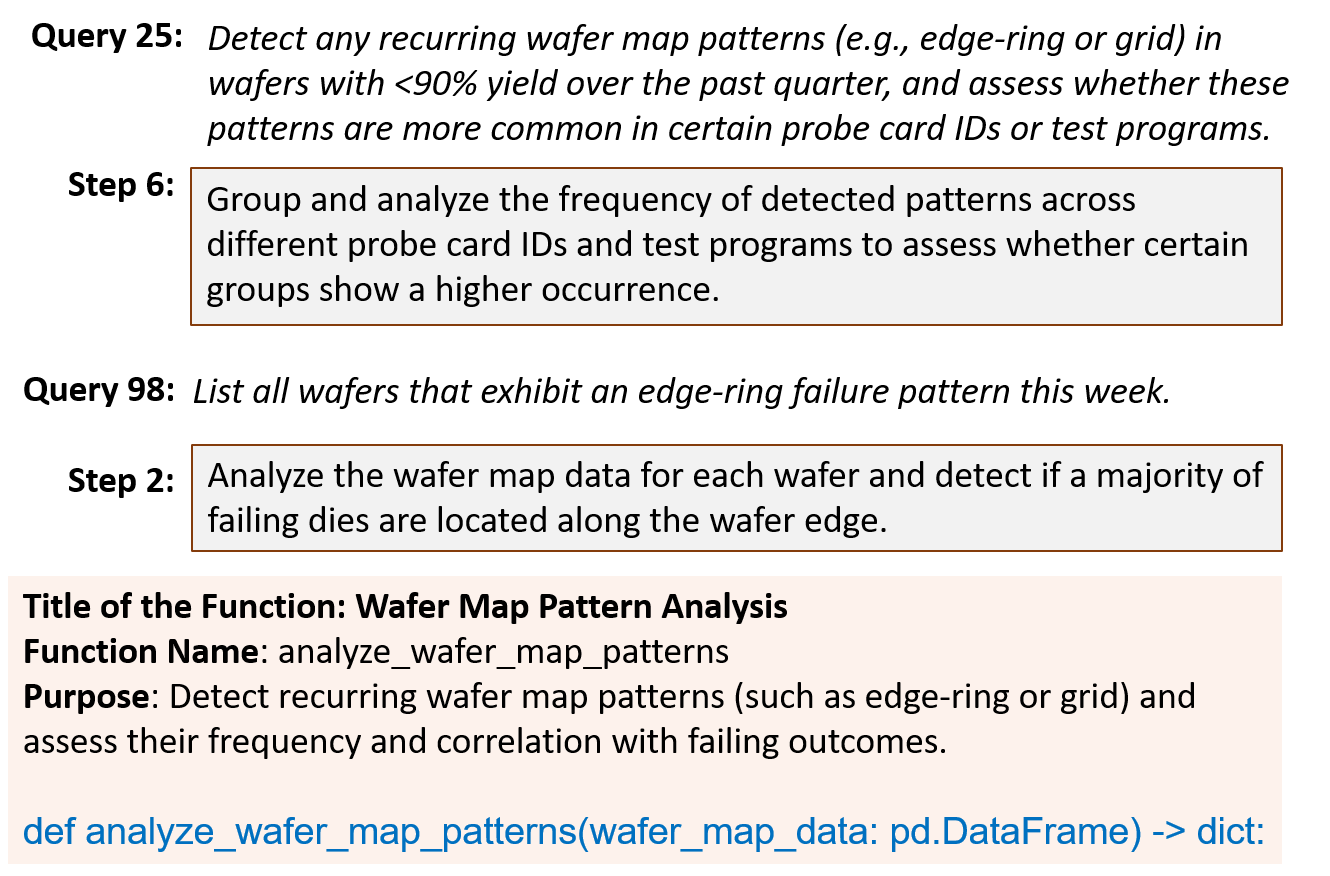}
	\vspace{-0.3cm}
	\caption{Function shared by two steps in the {\tt analyze} action group, one from the original 80 queries and the other from the extra 40 queries}
	\label{SharedF}
\end{figure}

Figure~\ref{SharedF} highlights a function frequently shared among steps within the {\tt analyze} action group -- the {\bf Wafer Map Pattern Analysis} function. The examples show that query 25 (from the original set) and query 98 (from the additional set) both require wafer map pattern analysis. In this case, the function initially prescribed by the model for query 25 was sufficiently general to also accommodate step 2 from the subsequently introduced query 98.

\begin{figure}[htb]
	\centering
	\vspace{-0.2cm}
	\includegraphics[width=3.3in]{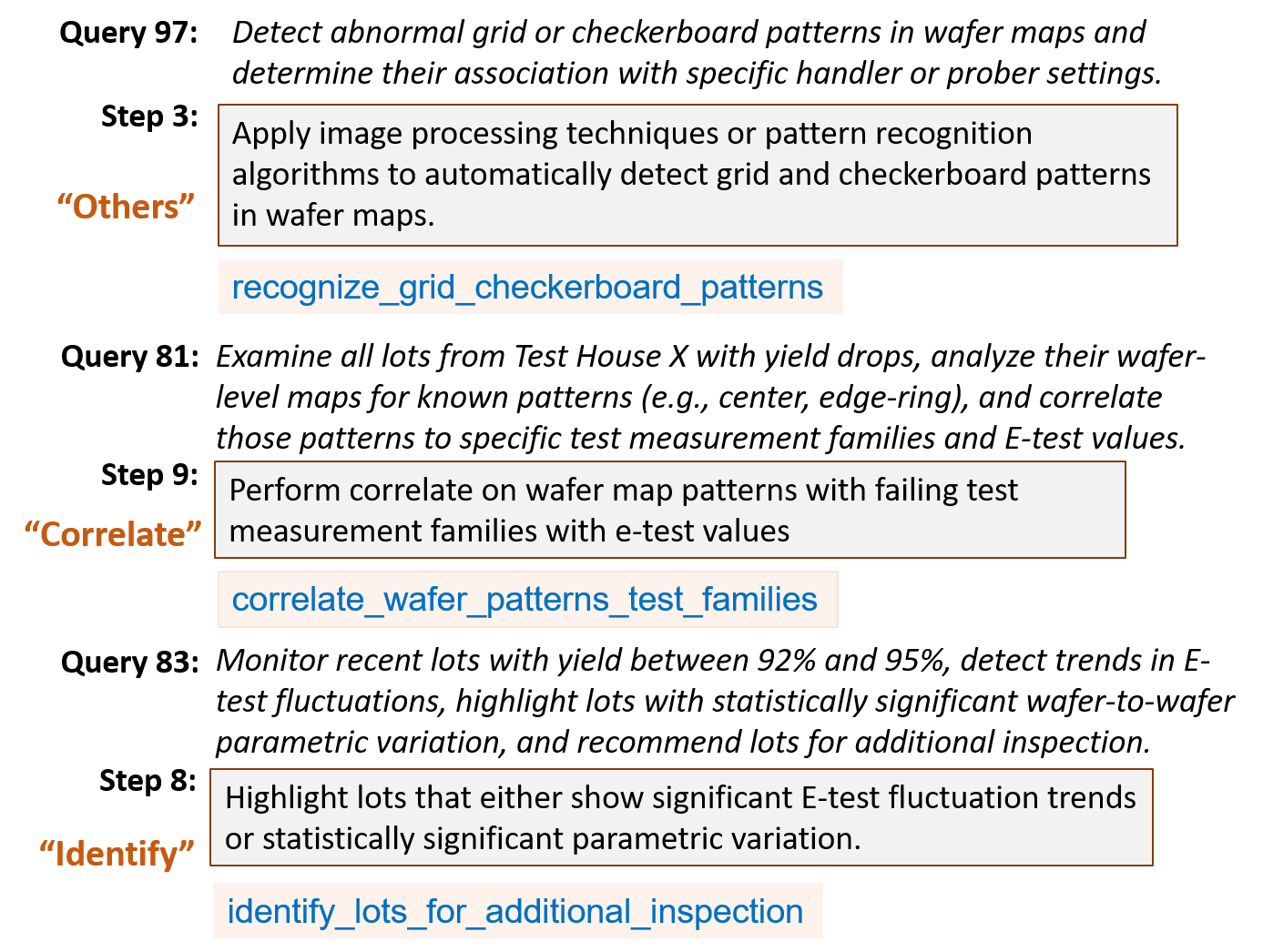}
	\vspace{-0.3cm}
	\caption{Unique functions created for steps from the additional 40 queries, in the actions groups {\tt Others}, {\tt Correlate}, and {\tt Identify}.}
	\label{UniqueF}
\end{figure}

In contrast, Figure~\ref{UniqueF} presents three examples that required the creation of unique functions tailored specifically to their respective steps. Step 3 from query 97 demands detecting grid and checkerboard patterns. The model intentionally distinguished this analysis from the earlier wafer map pattern analysis function. This indicates the model's robust understanding of wafer map analytics, as our experience confirmed that detecting grid and checkerboard patterns indeed necessitates specialized techniques distinct from those for other patterns (e.g., center, edge, ring) (see~\cite{ITC2023}).

Similarly, the function associated with step 9 from query 81 is unique, as it involves correlating wafer patterns with a family of E-test measurements -- a less common requirement that warrants specialized implementation. Step 8 from query 83 also possesses a distinctive characteristic, focusing on identifying significant fluctuations in E-test values across lots. These examples underscore that test data analytics encompasses a wide range of diverse perspectives, in many cases necessitating the inclusion of specialized functions.

\begin{table}[htb]
	\centering
	\setlength{\tabcolsep}{2pt}
	\begin{tabular}{cc|cc} \toprule
		\multicolumn{2}{c}{From original 80 queries} & \multicolumn{2}{c}{From 40 additional queries} \\ 
		{\bf \# of Steps} & {\bf \# of Functions} & {\bf Added \# of Steps} & {\bf Added \# of Functions} \\ \midrule
		135 &  37  & 63 & 16  \\ 
		\bottomrule
	\end{tabular}
	\vspace{0.1cm}
	\caption{{\bf Output:} \# of API functions automatically derived from the original 80 queries and additional 40 queries} 
	\label{ResultAPI}
	\vspace{-0.3cm}
\end{table}

\vspace{-0.4cm}
\subsubsection{Functions for the ``Output'' Category}
\label{sec04.2.2}

\begin{figure}[htb]
	\centering
	\vspace{-0.1cm}
	\includegraphics[width=2.4in]{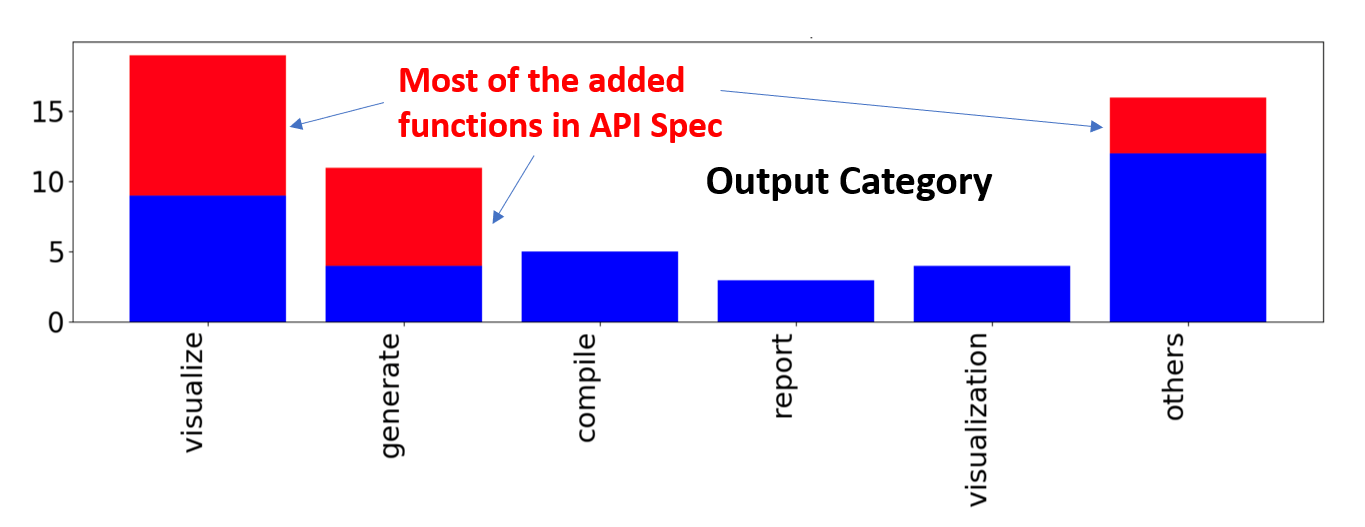}
	\vspace{-0.3cm}
	\caption{Histogram to show the \# of added functions in API Spec}
	\label{ResultH2}
	\vspace{-0.4cm}
\end{figure}

Parallel to Table~\ref{AnalysisAPI} and Figure~\ref{AnalysisH2}, Table~\ref{ResultAPI} and Figure~\ref{ResultH2} present corresponding results for steps in the Output category. Note from Table~\ref{ResultAPI} that the ratio from the original 80 queries ($\frac{37}{135}=27.4\%$) is comparable to that from the additional 40 queries ($\frac{16}{63}=25.4\%$); thus, we did not observe a significant drop in these ratios, unlike the situation previously observed with Table~\ref{AnalysisAPI}. However, these ratios are considerably smaller than those shown in Table~\ref{AnalysisAPI}, indicating that many more steps in the Output category share common functions. This aligns with our experience, as the variety of outputs (e.g., plot types) is generally more limited compared to the diverse range of analysis methods.

Figure~\ref{ResultH2} shows a similar characteristic observed before on Figure~\ref{AnalysisH2}, that reusable functions and those tailored to unique tasks are clearly separated into distinct groups by the model. This is a desirable property which will be elaborated further in Section~\ref{sec04.3}. 

\subsubsection{Data agent for the ``Data'' category}
\label{sec04.2.3}

As shown in Table~\ref{category}, 254 of the 588 steps and 133 of the 308 steps belong to the Data category. Instead of generating API specifications for them, we utilized a dedicated {\it Data} agent to automatically generate the necessary code. This became feasible recently due to LangGraph's support \cite{LangGraph_2024} for easy integration with popular database platforms such as Neo4j \cite{Neo4j}. Once the test data, such as wafer sort data, is organized into a Neo4j database, the LangGraph platform provides a template to import the Neo4j schema directly into an LLM prompt for query processing. The LLM then generates the required code to execute database operations specified by the query. Such operations typically include data retrieval, filtering, selection, and calculations for statistical reporting. Steps classified into the Data category can thus be effectively addressed through these database operations, resulting in the creation of a dedicated {\it Data} agent.

\begin{figure}[htb]
	\centering
	\vspace{-0.2cm}
	\includegraphics[width=2.4in]{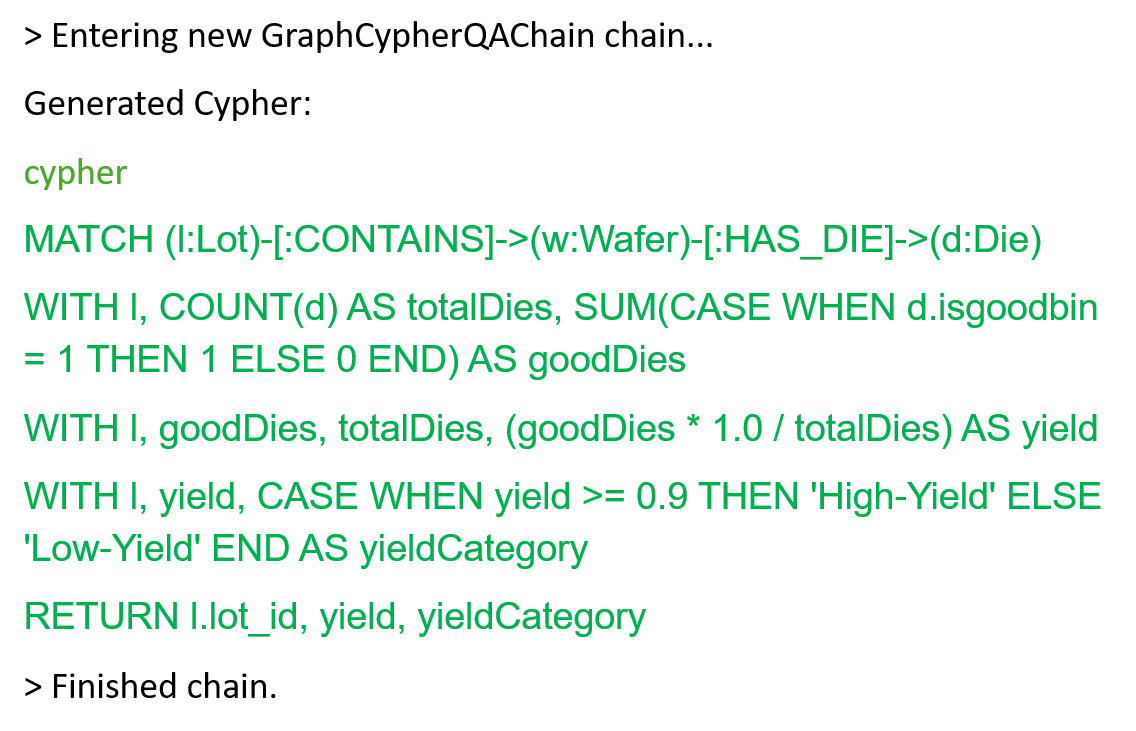}
	\vspace{-0.3cm}
	\caption{Cypher code generated by GPT-4o for Neo4j database, to perform Step 2 in Figure~\ref{ComplexQ} where high-yield is defined as $\geq 0.9$}
	\label{cypher}
	\vspace{-0.2cm}
\end{figure}

For instance, after we organized the wafer sort data used in \cite{ITC2022} into a Neo4j graph database and created a corresponding {\it Data} agent, the agent automatically generated the Cypher code shown in Figure~\ref{cypher} to perform Step 2 in Figure~\ref{ComplexQ}, which was categorized under Data category. Note that Cypher is a query language specifically for Neo4j.

\subsection{Addressing the scalability concern}
\vspace{-0.2cm}
\label{sec04.3}

Leveraging the capabilities of GPT-4o and GPT-o3-mini-high, we developed an automated approach to generate API specifications for steps categorized under {\bf Analysis} and {\bf Output}, while implementing a dedicated {\it Data} agent to handle steps in the {\bf Data} category. With these capabilities, we go back and ask the question: ``Does IEA-Plugin effectively address the scalability concern raised in Section~\ref{sec01.1}?'' 

It is natural to initially consider scalability in the following manner: Suppose our backend API comprises $n$ functions $f_1, \ldots, f_n$. When $k$ new requirements arise, the API needs to expand to accommodate $m$ functions, $f_1, \ldots, f_m$, with $m > n$. In this simplistic view, scalability could be measured by the ratio $\frac{m-n}{k}$. However, this perspective is no longer as relevant, given the robust code-generation capabilities offered by today's LLMs.

With LLMs, the overhead associated with coding new, well-defined functions is significantly reduced. Therefore, measuring overhead solely based on the number of functions to implement is no longer that meaningful. Instead, the primary challenge lies in structuring an API that remains stable even as new user requirements emerge. IEA-Plugin demonstrates that the true value provided by LLMs is precisely in {\it streamlining this API structure through a systematic analysis of user requirements}, represented by diverse queries.

The discussion in Section~\ref{sec04.2} highlights the capability of LLMs to automatically differentiate reusable functions from those tailored to unique tasks, a crucial property enabling the creation of a more stable API structure. As illustrated in Figures~\ref{AnalysisH2} and~\ref{ResultH2}, when new requirements arise, IEA-Plugin identifies where new functions should be integrated into the existing structure. Once they and their locations are clearly specified, the actual implementation effort becomes less of a scalability concern due to the efficient coding capabilities provided by LLMs. Consequently, IEA-Plugin effectively addresses scalability concerns by automatically generating a stable API structure, complemented by a dedicated Data agent capable of handling diverse data preparation tasks through automated database operations.   

\vspace{-0.1cm}
\section{Conclusion}
\vspace{-0.2cm}
\label{sec05}

While the development of IEA-Plugin was initially motivated by challenges encountered during the industrial deployment of IEA-Plot, the resulting IEA-Plugin was designed as an independent knowledge-acquisition tool. Rather than relying on meetings and documentation to acquire user requirements, IEA-Plugin enables knowledge acquisition directly through user interactions. Users simply enter their desired queries, confirm the automatically generated workflows, and IEA-Plugin enhances and assimilates this acquired knowledge into a structural API specification. 

The primary strength of IEA-Plugin lies in generating a systematic, stable API structure and clearly specifying the functions to be implemented, where their coding can be significantly facilitated by the capabilities of LLMs. In other words, the core value of IEA-Plugin is not in the coding itself, but in generating a structured plan to guide the coding process. Developing such a clear, systematic, and stable plan for test data analytics is especially challenging in companies facing a wide range of diverse requirements originating from various teams. This work demonstrates how such challenges can be effectively addressed by leveraging recent advancements in LLMs and AI-agent platforms.

IEA-Plugin employs three LLMs: GPT-4o for query generation and workflow summarization, GPT-o3-mini for workflow generation, and GPT-o3-mini-high for API specification generation. GPT-4o could also be integrated with the Neo4j platform to automatically generate Cypher code for database operations. Both the workflow reasoner and the dedicated {\it Data} agent were built upon the LangGraph platform. These technological advancements enabled the complete realization of IEA-Plugin in just under three months.

The most significant lesson we learned was that effectively leveraging the power of LLMs requires a shift in thinking from what we, as test practitioners, were previously accustomed to. Instead of insisting on fixed measures of quality, we learned to embrace and leverage variability. Likewise, rather than focusing on low-level coding tasks, we discovered the importance of emphasizing a structural, stable planning process to guide coding efforts. By introducing IEA-Plugin, we hope to share these valuable insights from our experiences with the test community.

\vspace{0.2cm}

\noindent
{\bf Acknowledgment}
The authors are thankful to Sergio Mier, Patty Pun and their teams at Qualcomm 
for the opportunity to deploy IEA and their valuable inputs to our research.

\bibliographystyle{IEEEtran}
\itemsep=0pt
\topsep = 0pt
\partopsep=0pt
\itemsep= 2pt
\parsep=0pt
\parskip= 0cm
\baselineskip=4pt

\input{Main.bbl}

\end{document}